\documentclass [12pt]{article}
\begin{document}

\begin{center}
\large\textbf{Hidden conformal symmetry of rotating charged black
holes}
\end{center}

\begin{center}
 Deyou Chen\footnote
{E-mail: deyouchen@uestc.edu.cn}, \quad Peng Wang \footnote {E-mail:
pengw@uestc.edu.cn}, \quad Houwen Wu \footnote {E-mail:
iverwu@uestc.edu.cn} and \quad Haitang Yang \footnote {E-mail:
hyanga@uestc.edu.cn}

\end{center}
\begin{center}
School of Physical Electronics, University of Electronic Science and
Technology of China, Chengdu,  Sichuan 610054, China
\end{center}

\textbf{Abstract:} Motivated by the recent work of the hidden
conformal symmetry of the Kerr black hole, we investigate the hidden
conformal symmetry of a Kerr-Sen black hole and a Kerr-Newman-Kasuya
black hole. Our result shows the conformal symmetry is spontaneously
broken due to the periodicity of the azimuthal angle. The absorption
across section is in consistence with the finite temperature
absorption cross section for a 2D CFT. The entropies of the black
holes are reproduced by the Cardy formula.

\textbf{1. Introduction}

The Kerr/CFT correspondence put forward by Guica, Hartman, Song and
Strominger \cite{GHSS} has attracted people's great interest. In
this work, quantum gravity in the region very near the horizon of an
extreme Kerr black hole was researched. After the specification of
boundary conditions at the asymptotic infinity of the near-horizon
extreme Kerr (NHEK) geometry and the demonstration of their
consistency, the authors showed the asymptotic symmetry group is one
copy of the conformal group and furthermore has a central charge
$c_L = 12J$. The macroscopic Bekenstein-Hawking entropy was
reproduced by the microscopical computation of entropy from the
Cardy formula. Subsequently, this work was extended to various
background spacetime and all of the work supports the Kerr/CFT
correspondence and give the evidences of holographic duality. The
related work can be seen in\cite{HHKNT}. In fact, the holographic
duality was early studied in the work of Brown and Henneaux \cite
{BH}.

Very recently, Castro and his collaborators \cite{CMS} have studied
the hidden conformal symmetry of the non-extreme Kerr background
spacetime by the massless scalar field and showed there is the
conformal symmetry, but it is spontaneously broken due to the
existence of $2\pi$ periodicity of the azimuthal angle. It should
point out that the conformal symmetry is not symmetry of the
spacetime geometry and it acts locally on the solution space. The
calculation of the absorption probability shows that the near region
of the Kerr background spacetime is dual to the 2D CFT. In the
microcosmic depiction, using the Cardy formula for the microstate
degeneracy, they reobtained the Bekenstein-Hawking entropy of the
black hole. Based on this work, Krishnan explored the hidden
conformal symmetry of the most general black holes in five
dimensions arising from heterotic/type II string theory
\cite{Krishnan}. Chen and Sun studied the hidden conformal symmetry
of the nonextremal uplifted 5D Reissner-Nordstrom black hole
\cite{CS}. The case of Kerr-Newman black holes was investigated by
Wang and Chen in \cite{WL} and \cite{CL}, respectively. Li et.al.
investigated the case of the Kaluza-Klein Black Hole \cite{LLR}.

Our aim in this paper is to explore the hidden conformal symmetry of
a Kerr-Sen black hole and a Kerr-Newman-Kasuya black hole by using
the idea of Castro and his collaborators. Our result is similar to
that of the Kerr black hole and the conformal symmetry is
spontaneously broken. The Bekenstein-Hawking entropies of the black
holes are recovered by the Cardy formula and the CFT temperatures
are obtained respectively. Finally, we investigate the absorption
across section of the Kerr-Sen black hole in the near region and get
the result that the near region of the Kerr-Sen black hole is dual
to 2D CFT. The similar case of the Kerr-Newman-Kasuya black hole is
investigated.

The rest of this paper is organized as follows. In sect. 2, the
hidden conformal symmetry of the Kerr-Sen black hole is explored by
the massless scalar field. In sect. 3, we explore the hidden
conformal symmetry of the Kerr-Newman-Kasuya black hole. Meanwhile,
the entropy and the absorption cross section are investigated. Sect. 4
includes our conclusions.

\textbf{2. Hidden Conformal Symmetry of the Kerr-Sen Black Hole}

The Kerr-Sen solution was derived from the low energy effective
action for heterotic string  theory \cite{SEN}. In the generalized
Boyer-Linquist coordinates, the metric is given by

\begin{equation}
ds^2 = - \frac{\Delta }{\Sigma }\left( {dt - a\sin ^2\theta d\phi
} \right)^2 + \frac{\Sigma }{\Delta }dr^2 + \Sigma d\theta ^2 +
\frac{\sin ^2\theta }{\Sigma }\left[ {adt - \left( {\Sigma +
a^2\sin ^2\theta } \right)d\phi } \right]^2,
\end{equation}

\noindent
with the electromagnetic potential

\begin{equation}
A_\mu = \frac{Qr}{\sqrt 2 \Sigma }\left( {dt - a\sin ^2\theta
d\phi } \right),
\end{equation}

\noindent where

\[\Delta = r^2 - 2{M}'r + a^2 = \left( {r - r_ + } \right)\left( {r -
r_ - } \right),
\]
\begin{equation}
 \quad \Sigma = r^2 + 2br + a^2\cos ^2\theta ,  r_\pm =
{M}'\pm \sqrt {{M}'^2 - a^2} ,
\end{equation}

\noindent $r_\pm $ are the outer (inner) horizons and ${M}' = M - b
= M - \frac{Q^2}{2M}$. $M$, $Q$ and $J = Ma$ are  physical mass,
charge and angular momentum of the black hole, respectively. The
horizon area, entropy, Hawking temperature and angular velocity at
the event horizon are

\[
A = 4\pi \left( {r_ + ^2 + 2br_ + + a^2} \right) = 8\pi Mr_ + ,
\]

\begin{equation}
S = 2\pi Mr_ + , \quad T = \frac{r_ + - {M}'}{4\pi Mr_ + }, \quad
\Omega _ + = \frac{a}{2Mr_ + }.
\end{equation}

\noindent In this paper, we only consider the massless scalar
particles. The solution of the Klein-Gordon equation for the
massless scalar field on the Kerr-Sen background spacetime is
obtained as follows

\[
\partial _r \left( {\Delta \partial _r \Phi } \right) - \frac{1}{\Delta
}\left[ {\left( {r^2 + 2br + a^2} \right)\partial _t + a\partial
_\phi } \right]^2\Phi + \frac{1}{\sin \theta }\partial _\theta
\left( {\sin \theta
\partial _\theta \Phi } \right)
\]

\begin{equation}
+ \left( {a\sin \theta \partial _t + \frac{1}{\sin \theta }\partial
_\phi } \right)^2\Phi = 0.
\end{equation}

\noindent The existence of two killing vectors $\partial_t$ and
$\partial_\phi$ enables us to separate the variables as $\Phi =
e^{ - i\left( {\omega t - m\phi } \right)}R\left( r \right)\Theta
\left( \theta \right)$, where $\omega$ and $m$ are quantum
numbers. Then the wave equation can be decomposed into two parts
(angular part and radial part), which are

\begin{equation}
 - \Lambda
\Theta =\frac{1}{\sin \theta }\partial _\theta \left( {\sin \theta
\partial _\theta \Theta } \right) - \left( {\omega ^2a\sin
^2\theta + \frac{m^2}{\sin ^2\theta }\partial _\phi }
\right)\Theta,
\end{equation}

\begin{eqnarray}
\Lambda R&=&\partial _r \left( {\Delta \partial _r R} \right) +
\frac{\left( {ma - 2Mr_ + \omega } \right)^2}{\left( {r - r_ + }
\right)\left( {r_ + - r_ - } \right)}R - \frac{\left( {ma - 2Mr_ -
\omega } \right)^2}{\left( {r - r_ - } \right)\left( {r_ + - r_ -
} \right)}R \nonumber\\
&&+ \left[ {\Delta + 4M\left( {M + r} \right)} \right]\omega ^2R.
\end{eqnarray}

\noindent We are interested in the case of the low frequency limit
$ \omega M\ll 1 $. In this case the geometry can be divided into
two parts and the terms included $\omega^2$ can be neglected in
eqs. (6) and (7). In the near region defined as $r\omega \ll 1$
\cite{CMS}, eqs. (6) and (7) are simplified as

\begin{equation}
 - \Lambda \Theta =\frac{1}{\sin \theta }\partial _\theta \left( {\sin \theta
\partial _\theta \Theta } \right) - \frac{m^2}{\sin ^2\theta
}\partial _\phi \Theta ,
\end{equation}

\begin{equation}
 \Lambda R=\partial _r \left( {\Delta \partial _r R} \right) + \frac{\left(
{ma - 2Mr_ + \omega } \right)^2}{\left( {r - r_ + } \right)\left(
{r_ + - r_ - } \right)}R - \frac{\left( {ma - 2Mr_ - \omega }
\right)^2}{\left( {r - r_ - } \right)\left( {r_ + - r_ - }
\right)}R .
\end{equation}

\noindent It is obvious to find that eq. (8) denotes the standard
Laplacian on the $2$-sphere. Thus one can get $\Lambda=l(l+1)$.
Comparing eq. (9) with the Kerr case, we find they have the similar
form, the difference is the values of $r_\pm$.

In fact, Eq. (9) can also be derived from the $SL(2,R)$ Casimir
operator, which reflects the $SL(2,R)_L \times SL(2,R)_R$ symmetry
of the near-region scalar field equation. This is improved in the
following. We follow Ref.\cite{CMS,Krishnan,CS,WL,CL,LLR} and
firstly introduce the conformal coordinates $(\omega^\pm, y)$ that
relate to the coordinates $(t, r, \phi)$ by

\[
\omega ^ + = \sqrt {\frac{r - r_ + }{r - r_ - }} e^{2\pi T_R \phi +
2n_R t},
\]
\[
\quad \omega ^ - = \sqrt {\frac{r - r_ + }{r - r_ - }}
e^{2\pi T_L \phi + 2n_L t},
\]
\begin{equation}
\quad y = \sqrt {\frac{r_ + - r_ - }{r - r_ - }} e^{\pi \left( {T_R
+ T_L } \right)\phi + \left( {n_R + n_L } \right)t}.
\end{equation}

\noindent The definition of the local vector field is

\begin{equation}
H_1 = i\partial _ + , \quad H_0 = i\left( {\omega ^ + \partial _ + +
\frac{1}{2}y\partial _y } \right), \quad H_{ - 1} = i\left[ {\left(
{\omega ^ + } \right)^2\partial _ + + \omega ^ + y\partial _y -
y^2\partial _ - } \right],
\end{equation}

\begin{equation}
\tilde {H}_1 = i\partial _ - , \quad \tilde {H}_0 = i\left( {\omega
^ - \partial _ - + \frac{1}{2}y\partial _y } \right), \quad \tilde
{H}_{ - 1} = i\left[ {\left( {\omega ^ - } \right)^2\partial _ - +
\omega ^ - y\partial _y - y^2\partial _ + } \right],
\end{equation}

\noindent which obey the $SL\left( {2,R} \right)$ Lie bracket
algebra

\begin{equation}
\left[ {H_0 ,H_{\pm 1} } \right] = \mp iH_{\pm 1} , \quad \left[
{H_{ - 1} ,H_1 } \right] = - 2iH_0 ,
\end{equation}

\noindent and the same commutators are hold for $\left( {\tilde
H_0 ,\tilde H_{\pm 1} } \right)$. The corresponding quadratic
Casimir is

\begin{equation}
{\rm H}^2 = \tilde {{\rm H}}^2 = - H_0^2 + \frac{1}{2}\left( {H_1
H_{ - 1} + H_{ - 1} H_1 } \right) = \frac{1}{4}\left( {y^2\partial
_y^2 - y\partial _y } \right) + y^2\partial _ + \partial _ - .
\end{equation}

\noindent To study the Casimir, we choose the identifications,

\[
\quad n_L = - \frac{1}{4M}, \quad T_L = \frac{r_ + + r_ - }{4\pi a}=
\frac{2M^2 - Q^2}{4\pi Ma},
\]

\begin{equation}
n_R = 0, \quad T_R = \frac{r_ + - r_ - }{4\pi a} =  \frac{\sqrt
{\left( {2M^2 - Q^2} \right)^2 - 4M^2a^2} }{4\pi Ma}.
\end{equation}

\noindent From eqs. (10), (14) and (15), the Casimir in term
of $(t, r, \phi)$ becomes

\begin{equation}
{\rm H}^2 = \partial _r \left( {\Delta \partial _r } \right) -
\frac{\left( {a\partial _\phi + 2Mr_ + \partial _t }
\right)^2}{\left( {r - r_ + } \right)\left( {r_ + - r_ - } \right)}
+ \frac{\left( {a\partial _\phi + 2Mr_ - \partial _t }
\right)^2}{\left( {r - r_ - } \right)\left( {r_ + - r_ - } \right)}.
\end{equation}

\noindent Comparing eq. (9) with eq. (16), we find the wave of
radial part in the near region can be rewritten as

\begin{equation}
{\rm H}^2\Phi = \tilde {{\rm H}}^2\Phi = l\left( {l + 1}
\right)\Phi,
\end{equation}

\noindent which shows the scalar Laplacian can be reduced to the
$SL(2,R)$ Casimir and the $SL(2,R)_L \times SL(2,R)_R$ weights of
the field $\Phi$ are $(l,l)$. We know the vector fields defined in
eqs. (11) and (12) are local and the angular $\phi$ is periodic.
Therefore they are not periodic under the angular identification $
\phi \sim \phi +2\pi$. This implies that the $SL(2,R)_L \times
SL(2,R)_R$ symmetry is spontaneously broken under the periodic
identification of the angular coordinate $\phi$.

In the following, we reproduce the Bekenstein-Hawking entropy of the
Kerr-Sen black hole from the Cardy formula for the dual 2D CFT.
There are many papers about the black hole entropy. The attractor
mechanism plays an important role when the scalars are turned on
(black holes in string theory). The research on the attractor
mechanism and entropy of extremal black holes based on the entropy
function formalism has been discussed \cite{Sen0708}. Here, we don't
know the corresponding the central charge $C_L$ and $C_R$ of the
non-extreme Kerr-Sen black hole. In
Ref.\cite{CMS,Krishnan,CS,WL,CL,LLR} the central charges near the
extreme case are adopted and are regarded as that of the non-extreme
case. The reason is that there is a smooth limit from near-extremal
to extremal solution and probably the hidden conformal symmetry
connects smoothly to that of extreme limit \footnote {We are
grateful to the reviewer for his help on this point.}. Therefore we
can get the central charges from the case of the extreme Kerr-Sen
black hole, which is

\begin{equation}
C_L = C_R = 12J.
\end{equation}

\noindent From the Cardy formula for the microstate degeneracy, we
 get
\begin{equation}
S = \frac{\pi ^2}{3}\left( {C_L T_L + C_R T_R } \right) = 2\pi Mr_ +
= S_{BH} ,
\end{equation}

\noindent where eq. (15) is introduced to get the second equal sign
and which shows the Bekenstein-Hawking entropy can be derived from
the dual 2D CFT.

Now we solve Eq. (7). There is much work appeared to solve such
equation in the research on greybody factors of black holes at the
frequency $\omega M\ll 1$ [\cite{GBF} and the references therein].
The difference here is that our consideration is the solution in
the near region $\omega r\ll1$. To bring Eq. (7) in the form of a
known differential equation, we introduce the following
transformation \cite{CL,GBF}

\begin{equation}
f = \frac{r - r_ - }{r - r_ + }.
\end{equation}

\noindent And then eq.(7) becomes

\begin{equation}
f\left( {1 - f} \right)\frac{d^2R}{df^2} + \left( {1 - f}
\right)\frac{dR}{df} +  \left\{\frac{\left[ {ma - 2Mr_+ \omega }
\right]^2}{f\left( {r_ + - r_ - } \right)^2} - \frac{\left[ {ma -
2Mr_-\omega } \right]^2}{\left( {r_ + - r_ - } \right)^2} -
\frac{l\left( {l + 1} \right)}{1 - f} \right\}R = 0.
\end{equation}

\noindent It is not a simply work to solve the above equation. We
can follow the method given in \cite{GBF} and get the solution. So
the absorption cross section is

\[
P_{acs}\sim \left| A \right|^{ - 2} = \sinh \left( {\frac{\omega - m\Omega
}{2T}} \right)\left| {\Gamma \left( {l + 1 - i2M\omega } \right)}
\right|^2
\]

\begin{equation}
\cdot \left| {\Gamma \left( {l + 1 - i2\frac{M\omega \left( {r_ + +
r_ - } \right) - ma}{r_ + - r_ - }} \right)} \right|^2.
\end{equation}

\noindent We find it has similar form with that of the Kerr-Newman
case, but the values of $r_\pm $ are different. We show that this
absorption across section relates to a 2D CFT in the following. To
do this, we should first find the conjugate charges $\left( {E_L
,E_R } \right)$ corresponding  to the CFT temperatures given by eq.
(15), which satisfy the relation

\begin{equation}
\delta S = \frac{\delta E_L }{T_L } + \frac{\delta E_R }{T_R }.
\end{equation}

\noindent The solution of the conjugate charges is easily gotten as

\begin{equation}
\delta E_L = \frac{\delta M}{a}\left( {r_+ + r_-} \right) M, \quad
\delta E_R = \frac{\delta M}{a}\left( {r_+ + r_-} \right)M - \delta
J
\end{equation}

\noindent after the first law of thermodynamics $T_H \delta S =
\delta M - \Omega _H \delta J - \Phi_H Q$ was introduced. In
\cite{CL}, the conjugate charges was derived with the consideration
of electro-magnetic field, where $\Phi_H Q \ne 0$. Since the
massless scalar field is considered in this paper, we do not care
$\Phi_H Q $. Hence the left and right moving frequencies are

\begin{equation}
\omega _L = \frac{\omega }{a}\left( {r_+ + r_-} \right) M, \quad
\omega _R = \frac{\omega }{a}\left( {r_+ + r_-} \right)M - m.
\end{equation}

\noindent where $\delta M = \omega $ and $\delta J = m$ were used
here. Combining eqs.(15), (22), (25) and the weights $(l,l)$, we can
get the absorption cross section as follows

\begin{equation}
P\sim T_L^{2h_L - 1} T_R^{2h_R - 1} \sinh \left( {\frac{\omega _L
}{T_L } + \frac{\omega _R }{T_R }} \right)\left| {\Gamma \left( {h_L
+ i\frac{\omega _L }{2\pi T_L }} \right)} \right|^2 \cdot \left|
{\Gamma \left( {h_R + i\frac{\omega _R }{2\pi T_R }} \right)}
\right|^2,
\end{equation}

\noindent which is just the finite-temperature absorption cross
section for a 2D CFT and shows the near region of the
Kerr-Sen background spacetime is dual to 2D CFT.

\textbf{3. Hidden Conformal Symmetry of a Kerr-Newman-Kasuya black
hole}

The hidden conformal symmetry of the charged rotating black hole has
been investigated \cite{WL,CL}. In this section, we investigate the
hidden conformal symmetry of a rotating black hole with electric
charge and magnetic charge (namely a Kerr-Newman-Kasuya black
hole) \cite{Kasuya}. The metric in the Boyer-Lindquist coordinates is given by

\[
ds^2 = - \left( {1 - \frac{2Mr - Q^2 - P^2}{\sum }} \right)dt^2 -
\frac{2a\sin ^2\theta }{\sum }\left( {2Mr - Q^2 - P^2} \right)dtd\varphi
\]

\begin{equation}
 + \frac{\sum }{\Delta }dr^2 + \sum d\theta ^2 + \left( {\frac{\left( {2Mr -
Q^2 - P^2} \right)a^2\sin ^2\theta }{\sum } - \left( {r^2 + a^2} \right)}
\right)\sin ^2\theta d\varphi ^2,
\end{equation}

\noindent with the electro-magnetic vector potential

\begin{equation}
A = \frac{Qr}{\sum }\left( {dt - a\sin ^2\theta d\phi } \right) +
\frac{P\cos \theta }{\sum }\left[ {adt - (r^2 + a^2)d\phi } \right]
+ \varepsilon Pd\phi ,
\end{equation}

\noindent where

\[
\Delta = r^2  - 2Mr+ a^2 + Q^2 + P^2 = \left( {r - r_ + }
\right)\left( {r - r_ - } \right),
\]

\begin{equation}
\sum = r^2 + a^2\cos ^2\theta , \quad \varepsilon = \pm 1,0,  \quad
r_\pm = M\pm \sqrt {M^2 - a^2 - Q^2 - P^2}.
\end{equation}

\noindent parameters $M$, $Q$ and $P$ are the physical mass,
electric charge and magnetic charge of the black hole, respectively.
The horizon area, entropy, Hawking temperature and angular velocity
at the event horizon are

\[
A = 4\pi \left( {2Mr_ + - Q^2 - P^2} \right), \quad S = \pi \left(
{2Mr_ + - Q^2 - P^2} \right),
\]

\begin{equation}
\quad T _H = \frac{r_ + - r_ - }{4\pi \left( {r_ + ^2 + a^2}
\right)}, \quad \Omega _H = \frac{a}{(r_ + ^2 + a^2)}.
\end{equation}

Since the Kerr-Newman-Kasuya metric can be obtained from
Kerr-Newman one by replacing $Q^2$ by $Q^2+P^2$, all the
calculations are parallel to those of Kerr-Newman, investigated by
\cite{CL}, we simply list the final conclusion here.
\begin{itemize}
\item Once considering a massless scalar field, the Laplacian can
be reduced to the $SL(2,R)$ Casimir in the near region $r\omega
\ll 1$.

\item With Cardy's formula and the available constraints $C_L =
C_R =12 J$, the CFT microstate degeneracy precisely coincides the
Bekenstein-Hawking entropy of the Kerr-Newman-Kasuya:

\begin{equation}
S = \frac{\pi ^2}{3}\left( {C_L T_L + C_R T_R } \right) = \pi
\left( {2Mr_ + - Q^2 - P^2} \right),
\end{equation}

\item In the near region, after imposing $r\gg M$, similar
calculation as in \cite{CL} shows that
\[
p_{abs} = \left| A \right|^{ - 2}\sim \sinh \frac{\omega - m\Omega
_H }{2T_H }\left| {\Gamma \left( {l + 1 - i2M\omega } \right)}
\right|^2
\]
\begin{equation}
\cdot \left| {\Gamma \left( {l + 1 + i2\frac{ma - \left( {2M^2 -
Q^2 - P^2} \right)\omega }{r_ + - r_ - }} \right)} \right|^2.
\end{equation}
With identifications of the corresponding left and right moving
frequencies

\[
\omega _L = \frac{2M^2 - Q^2 - P^2}{a}\omega ,
\]

\begin{equation}
\omega _R = \frac{2M^2 - Q^2 - P^2}{a}\omega - m,
\end{equation}
the two dimensional CFT finite-temperature absorption cross
section is reproduced:
\begin{equation}
p_{abs} \sim T_L^{2h_L - 1} T_R^{2h_R - 1} \sinh \left(
{\frac{\omega _L }{2T_L } + \frac{\omega _R }{2T_R }} \right)\left|
{\Gamma \left( {h_L + i\frac{\omega _L }{2\pi T_L }} \right)}
\right|^2 \cdot \left| {\Gamma \left( {h_R + i\frac{\omega _R }{2\pi
T_R }} \right)} \right|^2,
\end{equation}

\end{itemize}

\textbf{4. Conclusions}

In this paper, we have investigated the the dual conformal field
theory of the Kerr-Sen black hole and the Kerr-Newman-Kasuya black
hole. The result shows there is spontaneous breaking of the
conformal symmetry due to the existence of the periodic
identification of the azimuthal angle $\phi$ in the background
spacetime of the black holes. For the Kerr-Sen black hole, we get
the CFT temperatures as $T_L = \frac{2M^2 - Q^2}{4\pi Ma}$ and $T_R
= \frac{\sqrt {\left( {2M^2 - Q^2} \right)^2 - 4M^2a^2} }{4\pi Ma}$.
For the Kerr-Newman-Kasuya black hole, the CFT temperatures are $T_L
= \frac{2M^2 - Q^2-P^2}{4\pi Ma}$ and $T_R = \frac{\sqrt
{M^2-Q^2-P^2-a^2} }{2\pi a}$. Meanwhile the Bekenstein-Hawking
entropies of the black holes were reobtained by the Cardy formula.
When the conjugate charges and the CFT temperatures were introduced,
we find the absorption cross section of the Kerr-Sen black hole is
just the finite temperature absorption cross section for a 2D CFT.
There is the same case for the Kerr-Newman-Kasuya black hole. This
shows the absorption across section in gravity is in consistency
with that in CFT and the near region of the background spacetime of
the black hole is dual to 2D CFT.

\section*{Acknowledgments}
One of the authors (D. Chen) would like to thank Alejandra Castro
and Chiang-Mei Chen for their help. This work is supported by
Fundamental Research Funds for the Central Universities (Grant No.
ZYGX2009J044, ZYGX2009X008),  NSFC (Grant No.10705008) and NCET.

\end{document}